\documentclass[aps,twocolumn,showpacs,prl]{revtex4}
\usepackage{graphics,epsfig}
\newcommand{\up}{\uparrow}
\newcommand{\down}{\downarrow}

\begin{document}
\title{Order-disorder transition in nanoscopic semiconductor quantum rings}
\author{Peter Borrmann}
\author{Jens Harting}
\affiliation{Department of Physics, Carl von Ossietzky University
Oldenburg, D-26111 Oldenburg, Germany}
\date{\today}
\begin{abstract}
Using the path integral Monte Carlo technique we show that
semiconductor quantum rings with up to six electrons exhibit a
temperature, ring diameter, and particle number dependent transition
between spin ordered and disordered Wigner crystals. Due to the
small number of particles the transition extends over a broad
temperature range and is clearly identifiable from the electron pair
correlation functions. 
\end{abstract}
\maketitle
\begin{figure}[h]
\centerline{\epsfig{file=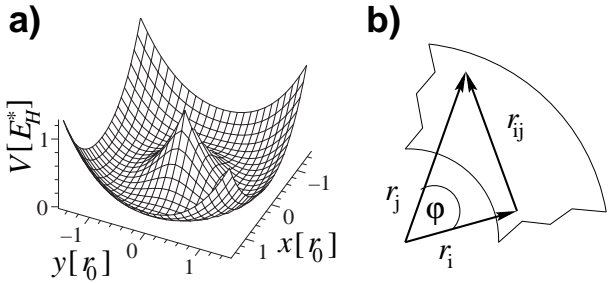,width=8cm}}
\caption{Potential energy surface of a semiconductor quantum ring (a) and
schematic illustration of the definition of $\varphi$ (b).}
\label{potfig}
\end{figure}

Nanoscopic semiconductor quantum rings (QRs), which recently have been
experimentally realized by Lorke {\it et.~al}~\cite{Lorke:2000}, are next to
quantum wires probably the best prototypes of quasi one dimensional
quantum systems. QRs can be viewed as rotating Wigner
crystals with promising features for application in microelectronics.
They can be modeled using a simple Hamiltonian of the form
\begin{equation}
H = \sum_{i=1}^{n} \left(
\frac{{\bf p}_i^2}{2 m^*} + \frac{1}{2} m^*\omega_0^2 (r_0-r_i)^2  \right)
+ \sum_{i<j} \frac{e^2}{\kappa r_{ij}},
\end{equation}
where $\kappa = 12.9$ and $m^* = 0.067 m_e$ are the material
constants of GaAs \cite{Hirose:1999}\footnote{The value of $\kappa$=12.9 
for GaAs is slightly different from 
the for the experiments of Lorke et al. more appropiate value of 
$\kappa$=14.5 for InAs. We have chosen this value for easier comparison 
to previous quantum dot calculations.}.  
The parameter $r_0$ is the radius of the quantum
ring and $\omega_0$ defines the strength of the two-dimensional
potential \cite{Lorke:2000,Chakraborty:1994}. Fig.~\ref{potfig}(a)
displays the shape of the ring potential. QRs can be tuned from 
quasi one-dimensional to two-dimensional systems by variation of 
the ring diameter and the potential strength. 

While mesoscopic QRs have been investigated theoretically and
experimentally in depth
\cite{Wendler:1995}, nanoscopic rings with strong quantum
effects are of increasing interest. Koskinnen
{\it et~al.}~\cite{Koskinen:2000} reported CI calculations of rotational
and vibrational excitations of nanoscopic QRs with up to
$N$=7 electrons. They claim that QRs behave like rather
rigid molecules or Wigner crystals with anti-ferromagnetic order in
the ground state. Ahn {\it et~al.}~\cite{Ahn:2000} considered stacked
nanoscopic rings and found an $N$- dependent Stark effect.
Experimentally it has been found that the emission energies of
QRs change abruptly whenever adding an electron
\cite{Warburton:2000}. 

In this letter we present the results of path integral Monte Carlo
(PIMC) simulations of single nanoscopic QRs with up to 8
electrons and different radii $r_0$.  We show that they undergo a
temperature, radius and particle number dependent spin
order-disorder transition. Furthermore, the influence of quantum
effects on the spatial electron distribution as well as the addition
energies $\Delta E$ are given.  Our results for the addition
energies, i.e.~the energy that is needed to place an additional
electron in a ring, is compared to the experimental results of
Warburton {\it et~al.}~\cite{Warburton:2000}.

In contrast to Hartree-Fock and spin density functional theory (DFT)
PIMC samples without any approximation the full many body wavefunction
instead of single or sums of Slater determinants. Especially for highly
correlated electron systems this is a major advantage of PIMC. Another
benefit of PIMC is the possibility to study temperature dependent
phenomena.
For quantum dots the problems of different 
density functional approaches have intensively been discussed in 
the past \cite{Hirose:1999,YL99,Harting:2000,Koskinen:1997}.
 The so called fermion sign problem is still a
topic of actual research and restricts the application of PIMC to a
limited number of fermions and for QRs to a temperature of at least 10
K.

\begin{figure*}[thbp]
\centerline{\epsfig{file=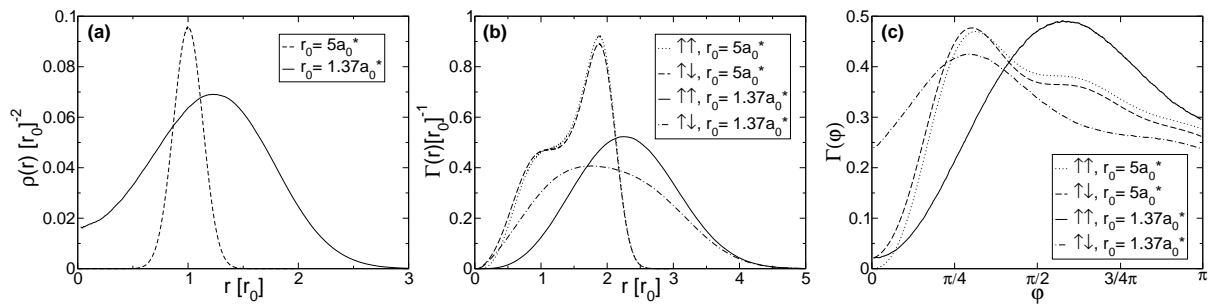,width=16cm}}
\caption{Radial electron density (a), radial (b) and angular (c)
pair correlation functions for $N$=6, $S$=0, and $r_0=1.37\,a_0^*$
and $5\,a_0^*$ at $T=15$ K.}
\label{r0G}
\end{figure*}
The Feynman path integral for an $N$-electron system with position
eigenket ${\mid {\bf r}_i,s_i\rangle}$ ($s_i =\pm \frac{1}{2}$ for
spin-up and spin-down electrons) in an external potential can be
rewritten as \cite{Takahashi:1984a}
\begin{eqnarray} \label{pfad}
Z &=& \int
\left[
\prod_{\gamma=1}^{M} \prod_{i=1}^{N} {\rm d}{\bf r}_i(\gamma)
\right]
\prod_{\delta=1}^{M} \det(A(\delta,\delta+1))\\
&\times&\exp\left(-\frac{\beta}{M} \sum_{\alpha=1}^{M}
V({\bf r}_1(\alpha),\ldots,{\bf r}_N(\alpha))\right) +
{\cal O}\left(\frac{\beta^3}{M^2}\right)\nonumber
\end{eqnarray}
with
\begin{eqnarray}
&&(A(\alpha,\alpha+1))_{i,j} \\
&&=\left\{
\begin{array} {l@{\quad:\quad}r}
\langle {\bf r}_i(\alpha)\mid
\exp\left( -\frac{\beta}{M} \frac{{\bf p}^2}{2m} \right) \mid {\bf
  r}_j
(\alpha+1)\rangle & s_i = s_j\\
0       & s_j \neq s_j
\end{array}
\right.\nonumber
\end{eqnarray}
and the boundary condition ${\bf r}_j(M+1)={\bf r}_j(1)$.  For 
$M \to \infty$ eq.~(\ref{pfad}) becomes exact.
Standard Metropolis Monte Carlo (MC) techniques can be utilized to
evaluate the integral in (\ref{pfad}).

The basic quantities reflecting the spatial structure of the electron
configuration are the electron-electron (distance) pair correlation
functions $\Gamma_{i,j}(r)=\langle\delta(r-$$\mid$${\bf
r}_{ij}\mid$$)\rangle$, the angular pair correlation functions
$\Gamma_{i,j}(\varphi)= 
\langle\delta(\varphi-$$\mid$$\varphi_i-\varphi_j$$\mid)\rangle$, 
and the radial electron density
$\rho_{i}(r) = \frac{1}{2 \pi r} \langle \delta( r - \mid {\bf r}_i\mid)
\rangle$, from which all energies can be calculated using the
Hypervirial theorem (for details of our method see \cite{Harting:2000}).
The definitions of $\varphi$ and ${\bf r}_{ij}$ are illustrated in
Fig.~\ref{potfig}(b).  To take the particle symmetry into account
we introduce
$\Gamma_{ij}=\Gamma^{\up\down}$ for $s_i\neq s_j$,
$\Gamma_{ij}=\Gamma^{\up\up}$ for $s_i= s_j=\frac{1}{2}$, and
$\Gamma_{ij}=\Gamma^{\down\down}$ for $s_i= s_j=-\frac{1}{2}$,
respectively. Obviously for $S=0$ we have $\Gamma^{\down\down}=
\Gamma^{\up\up}$.

In our simulations we controlled the systmatic error arising from the
limited number of timeslices $M$ and the statistical MC error carefully.
By choosing $M\times T\simeq 600$ and using up to 10 billion (10$^{10}$)
MC steps per run we pushed the overall error of all energy expectation
values below 0.3 $\%$.  Our Fortran-code is completely parallelized
using MPI and Lapack and most calculations have been performed on a Cray
T3E with 64 processors.

We fixed the strength of the harmonic potential $\hbar\omega_0=12$ meV,
resulting in effective atomic units for the Hartree energy
$E_H^*=10.995$ meV and the Bohr radius $a_0^*=10.1886$ nm. 
A ring diameter $r_0=14$ nm $= 1.37\,a_0^*$ than corresponds 
to the experimental setup of Lorke {\it et~al.}. To investigate 
the ring size dependence we performed additional calculations for 
$r_0=50.94$ nm = $5\,a_0^*$.

Fig.~\ref{r0G}(b) and (c) display the pair correlation functions of
QRs with $N=6$, $S=0$ and ring diameters $r_0=1.37\,a_0^*$
and $5.0\,a_0^*$ at $T=15$ K.  As expected for both diameters peaks at
$\varphi \approx \pi/3 $ and $\varphi \approx 2 \pi/3$ occur in
$\Gamma(\varphi)$. For $r_0 = 5\,a_0^*$ the angular pair correlation
functions for electrons with equal and unequal spin are almost
identical, i.e.~the Pauli principle does not play an important role in
this case. In contrast, for $r_0 = 1.37\,a_0^*$ the pair correlation
functions show a strong spin dependence. The electrons arrange on
the ring with anti-ferromagnetic order. Such spin density waves have been
predicted by Koskinen et al.~\cite{Koskinen:1997,Koskinen:2000}. 
The role of quantum effects
is reflected as well in Fig.~\ref{r0G}~(a). For the smaller ring
size the radial electron density is much broader and nonvanishing at
the ring center, implying that the system does not behave like a
quasi 1D-system. In a perfect hexagonal Wigner crystal the
equilibrium distances of the electrons would be $r$=1, 1.73, and
2.0$r_0$. For $r_0= 5.0\,a_0^*$ the distance pair correlation function
is approximately a properly weighted superposition of Gaussians
centered at these distances. For $r_0 =1.37\,a_0^*$ the pair
correlation function depends on the total spin of a pair and is
broadened up to 4$r_0$. From Fig.~\ref{r0G}(b) and (c) we infer that
the most probable configuration is one where the electrons are
ordered on a zig-zag line around the circle, i.e.~the electrons
arrange alternately in the inner and the outer part of the ring.

\begin{figure}[h]
\centerline{\epsfig{file=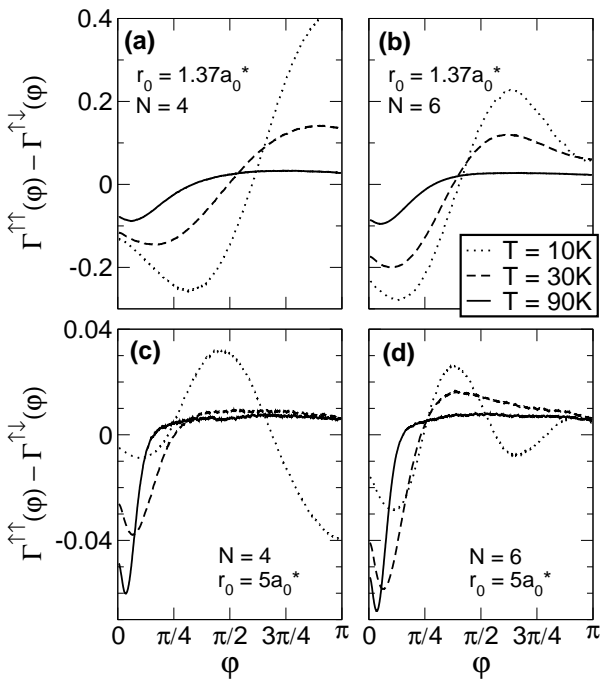,width=8cm}}
\caption{Angular spin density correlation function
($\Gamma^{\up\up}(\varphi) -\Gamma^{\up\down}(\varphi)$) for $T=10$,
30, and 90~K and 
(a) $N=4$, $S=0$, and $r_0 = 1.37 a_{0}^{*}$,
(b) $N=6$, $S=0$, and $r_0 = 1.37 a_{0}^{*}$,
(c) $N=4$, $S=0$, and $r_0 = 5 a_{0}^{*}$, and
(c) $N=6$, $S=0$, and $r_0 = 5 a_{0}^{*}$.}
\label{pauli}
\end{figure}

Fig.~\ref{pauli} displays the angular spin density pair correlation
functions for both ring diameters, temperatures $T$=15, 30, and 90 K
and particle numbers $N$=4, and 6. In all cases the large angle
correlations disappear with increasing temperature and at $T=90$ K
only at small angles a spin correlation is still visible. For $r_0 =
5.0\,a_0^*$ the negative correlation at small angles increases with
increasing temperature. However, this is simply due to the fact that
the values of both correlation functions at small angles become
larger with increasing temperature. The spin correlations for the
small ring are about one order of magnitude larger than those in the
ring with radius $r_0 = 5.0 a_{0}^{*}$ (Note the different scalings
of the abscissae). A comparison between $N=4$ and $N=6$ shows that
the spin correlation is smaller for the larger system. A probable
reason for this is that for $N=6$ the contribution of the Coulomb
repulsion is much larger (see Table~\ref{tab1}) freezing the electron in the Wigner crystal
and -- thinking in the picture of one-particle wavefunctions --
making the overlap between one-particle wavefunctions smaller.

In summary, from Fig.~\ref{pauli} can be inferred that a spin
order-disorder transition appears with increasing temperature,
increasing electron number and increasing ring size.

\begin{figure}[h]
\centerline{\epsfig{file=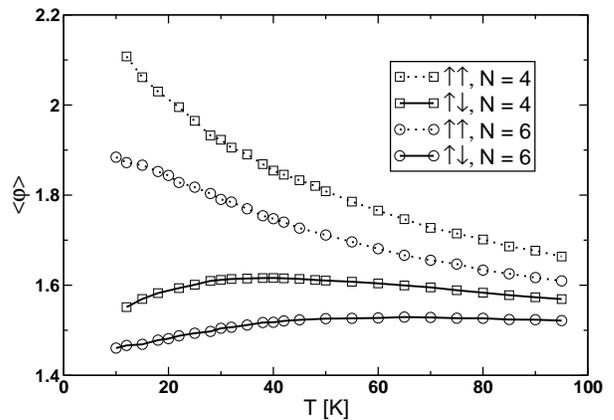,width=8cm}}
\caption{Temperature dependence of the mean angles 
$\langle~\!\!\phi^{\up\up}\rangle$ and $\langle\phi^{\up\down}\rangle$
for $N$=4 and 6, $S$=0 and $r_0=1.37\,a_0^*$.}
\label{PhiUUUD}
\end{figure}

Next we consider the temperature dependence of the spin order-disorder
transition in some more detail. Fig.~\ref{PhiUUUD} displays the mean
values of the angular separation of the electrons with equal and unequal
spin for $N=4$ and 6, $S=0$, and $r_0 = 1.37\,a_0^*$ as a function of
temperature. As expected from the results presented above the overall
differences between the expectation values for equal and unequal spins
approach zero with increasing temperature, i.e. the Pauli principle
becomes less important. In addition, the values for $N=6$ are smaller
than those for $N=4$. Obviously, this is because the available portion 
per particle of the ring volume is smaller for a larger number of
electrons and the Coulomb repulsion is unable to disperse the particles.
At $T= 10$ K and $N=4$ the  contribution of the Coulomb term to the
total potential energy is 53 $\%$ for $r_0 = 1.37\,a_0^*$ and only 36
$\%$ for $r_0 = 5\,a_0^*$, while for $N=6$ the difference between the 
different diameters is with 60 $\%$ and 51 $\%$ substantially smaller
(see Table~\ref{tab1}).

The slope of $\langle \varphi^{\up \down} \rangle$ for $N$=4 can be
understood as follows. Up to 40 K  $\langle\varphi^{\up \down}\rangle$
 increases due
spin disordering. At higher temperatures $\langle\varphi^{\up \down}\rangle$
decreases due to increasing thermal fluctuations. 

\begin{figure}[h]
\centerline{\epsfig{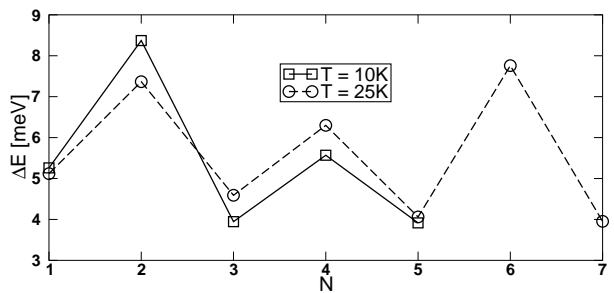}}
\caption{Addition energies $\Delta E$ at temperatures $T$=10 and 25~K
  for $r_0=1.37\,a_0^*$.}
\label{EADD}
\end{figure}

Finally, we calculated the second energy differences $\Delta E =
E_{N+1} - 2 E_{N} + E_{N-1}$ also called addition energies, which
are an indicator of the stability of a quantum ring with a given
number of electrons. For quantum dots it was claimed that the
electron configurations are given by Hunds rule \cite{Koskinen:1997}
and consequently magic numbers occur at $N=2,6,12$, and $20$. Here
we calculated only the addition energy up to $N=7$ for 25 K and $N=5$
for 10 K. As expected due to the Pauli principle and the Wigner
crystal structure a strong odd even effect occurs.  The general
behavior of this effect does not change for 25 K. However, for higher 
temperatures significant differences can be expected (see above).
Warburton {\it et~al.}~\cite{Warburton:2000} argued that
the general features of shell effects occurring in QRs are
the same as in quantum dots.  As can be inferred from
Fig.~\ref{EADD}, our calculations confirm this for $N=6$.
Furthermore, the addition energies are in the same range as those
from photoluminescence measurements at 4 K \cite{Warburton:2000}.
Due to the strong Coulomb repulsion in QRs, which grows
with increasing electron number, it can be expected that shell
effects become less important with increasing particle number.  

The effects described above are reflected quantitatively in Table
\ref{tab1}  presenting the total energy ($E_{\rm tot}$), kinetic energy
($E_{\rm kin}$), total potential energy  ($E_{\rm pot}$), the energy due
to the ring potential  ($E_{\rm ring}$), and the Coulomb energy ($E_{\rm
c}$). 
\begin{table}[h]
\caption{Total-, kinetic-, potential-, ring-, and Coulomb- energies for
  $N$=4 and 6, $S=0$, $r_0=1.37$ and 5 $a_0^*$ for different
  temperatures. All energies are given in meV.}
\begin{tabular}{cccp{0.8cm}p{1cm}p{1cm}p{1cm}p{1cm}p{1cm}}
\hline\hline
$N$&$r_0[a_0^*]$&$T$[K]&$E_{\rm tot}$&
$E_{\rm kin}$&$E_{\rm pot}$&$E_{\rm ring}$
&$E_{\rm c}$\\
\hline
4&1.37&10&57.2&23.0&34.2&16.1&18.1\\
 &    &25&59.3&24.4&35.0&16.8&18.2\\
 &    &90&85.1&40.6&44.4&26.5&17.9\\
\hline
 &5   &10&32.7&15.0&17.7&11.4&6.3\\
 &    &25&36.1&17.7&18.4&11.6&6.8\\
 &    &90&62.0&35.4&26.6&18.5&8.1\\
\hline\hline
6&1.37&10&117.8&39.4&78.4&31.1&47.3\\
 &    &25&122.5&42.6&79.9&32.9&46.9\\
 &    &90&160.9&66.7&94.2&48.4&45.8\\
\hline
 &5   &10&61.0&25.3&35.7&17.3&18.3\\
 &    &25&65.7&28.8&36.9&17.8&19.1\\
 &    &90&105.8&55.3&50.6&29.1&21.5\\
\hline\hline
\end{tabular}
\label{tab1}
\end{table}

In conclusion, we presented the results of full many body wavefunction
calculations for QRs with up to eight electrons.  We found that the
properties of the rings depend in an intriguing manner on the ring
diameter, the particle number and the temperature, which in turn is due
to spin correlation, Coulomb ordering and the general strength of
quantum effects. QRs exhibit a parameter dependent spin order-disorder
transition. By variation of the ring diameter the system can be tuned
from a quasi 1D Wigner crystal to a 2D structure.  The accesible
parameter ranges can be used to tune the properties of quantum rings to
desired values. Because of the ring diameter as an additional parameter
this qualifies them as even better candidates than quantum dots for
possible applications in microelectronics.  The addition energies
calculated using PIMC are in good agreement with the experimental
results of Warburton {\it et~al.}~\cite{Warburton:2000} and reflect the
predicted shell effects. 

We wish to thank the {\sl Konrad Zuse Institut Berlin} and the {\sl
Regionales Rechenzentrum Niedersachsen} for their excellent computer
support and Oliver M\"ulken, Heinrich Stamerjohanns and E.~R.~Hilf
for fruitful discussions.
%
%

\end{document}